\documentclass[epj]{svjour}
\usepackage{graphics}
\begin{document}
\title{Macroscopic and Local Magnetic Moments in Si-doped CuGeO$_3$
with Neutron and $\mu$SR Studies.}
\titlerunning{Macroscopic and local magnetic moments in Si-doped 
CuGeO$_3$}
%\subtitle{Do you have a subtitle?\\ If so, write it here}
\author{F.~Semadeni \inst{1}, 
A.~Amato\inst{2}\thanks{\emph{e-mail:} Alex.Amato@psi.ch},
B.~Roessli\inst{1}, P.~B\"oni\inst{3}, C.~Baines\inst{2},
T.~Masuda\inst{4}, K.~Uchinokura\inst{4} \and G.~Shirane\inst{5}  % etc
%\thanks{\emph{Present address:} Insert the address here if needed}%
}                     % Do not remove
%
%\offprints{}          % Insert a name or remove this line
%
\institute{Labor f\"ur Neutronenstreuung, ETH-Z\"urich \& PSI, 
  CH-5232 Villigen PSI, Switzerland
\and Laboratory for Muon-Spin Spectroscopy, Paul Scherrer Institute, 
CH-5232 Villigen PSI, Switzerland 
\and Physik Department E21, Technische Universit\"at M\"unchen, D-85747, 
Garching, Germany
\and Department of Advanced Materials Science, The University of Tokyo, 
Bunkyo-ku, Tokyo 113-8656, Japan 
\and Department of Physics, Brookhaven National Laboratory, 
Upton, New York 11973-5000, U.S.A.}
\date{Received: date / Revised version: date}
% The correct dates will be entered by Springer
%
\titlerunning{Macroscopic and Local Magnetic Moments in Si-doped CuGeO$_3$}
\authorrunning{F.~Semadeni et al.}
\abstract{
The temperature-concentration phase diagram of the Si-doped 
spin-Peierls compound CuGeO$_{3}$ is investigated by means 
of neutron scattering and muon spin rotation spectroscopy 
in order to determine the microscopic distribution 
of the magnetic and lattice dimerised regions as a function
of doping. 
The analysis of the zero-field muon spectra has
confirmed the spatial inhomogeneity of 
the staggered magnetisation that characterises the antiferromagnetic
superlattice peaks observed with neutrons.
In addition, the variation of the macroscopic order parameter with 
doping
can be understood by considering the evolution of the 
local magnetic moment as well as of the various
regions contributing to the muon signal.
 % end of PACS codes
} %end of abstract
\maketitle
\section{Introduction}
\label{intro}
Since the first observation that the inorganic compound CuGeO$_{3}$ 
undergoes a spin-Peierls (SP) transition \cite{Hase93}, 
an extensive study of the doped system has been undertaken. 
Site (Cu$_{1-x}$M$_{x}$GeO$_{3}$) \cite{Coad96,Hase96,Martin97}
and bond (CuGe$_{1-x}$Si$_{x}$O$_{3}$) doping 
\cite{Regnault95,Hirota98} 
studies have revealed the existence of a new antiferromagnetic phase 
below the SP ordering temperature T$_{SP}$.
Grenier et al. \cite{Grenier98-1} have shown that the
temperature concentration(T-x) phase diagram for site
and bond doping in CuGeO$_3$ exhibits a similar behaviour. The 
main features are the reduction of T$_{SP}$ with increasing doping
concentration, the suppression of the SP transition at a critical 
concentration x$_{c}$, the onset of an AF phase characterised by 
an ordering temperature T$_{N}$ increasing with x, and decreasing
for concentrations larger than x$_{c}$. The critical concentration 
x$_{c}$
is of the order of 3\% for site doping and about three times smaller 
for the bond doping scenario.

An inspection of Mg-doped compounds near x$_{c}$
by means of magnetic susceptibility \cite{Masuda98} and neutron 
scattering measurements \cite{Nakao99} has provided a more detailed
description of the phase diagram for the site-doped SP.
The ordering temperature T$_{N}$ and the order parameter
$\mu_{eff}$ of the antiferromagnetic phase exhibit a
discontinuity at x$_{c}$ $\approx$ 2.7\%.
The interpretation that has been proposed is that doped 
CuGeO$_{3}$ undergoes a first order phase transition at the 
critical concentration x$_{c}$ that constitutes a compositional 
phase boundary. A distinction is therefore introduced between
a dimerised AF phase (D-AF) below x$_{c}$ , and a uniform AF phase 
for 
higher concentration (U-AF).
Susceptibility measurements near x$_{c}$
performed by Masuda et al. \cite{Masuda00} have revealed the 
presence of a double peak region, which was interpreted as
a coexistence of both D-AF and U-AF phases.
 In addition they have 
performed high-resolution synchrotron diffraction on the (3/2 1 3/2)
superlattice reflection. For low doping, the X-ray peak profile is 
resolution 
limited, indicating the presence of long range order in the lattice 
dimerisation. In the region of doping which exhibits the  double peak in 
the
susceptibility, the superlattice X-ray peak broadens, thus indicating
the onset of short range order in the SP phase.

On the other hand, an extensive study of Si-doped compounds with 
susceptibility measurements was perfor\-med by Grenier et 
al. \cite{Grenier98-2}. They have shown that the 
introduction of Si ions reduces the intensity of 
the signal attributed to the dimerised phase. Moreover
an additional peak appears below T$_{SP}$, which 
is interpreted as the onset of an antiferromagnetic ordering
at a temperature T$_{N}$.
Above T$_{N}$, they observe an additional intensity in the 
susceptibility
which is attributed to the freeing of S=1/2 spins near the 
doping centers. The proportion of the SP signal linearly decreases
with increasing doping, whereas the proportion of free spins 
increases.
However, no double peak feature is
observed in the susceptibility measurements in the vicinity of the
critical doping concentration where the SP phase collapses.

In order to explain the effect of impurities on the SP phase
a theoretical model has been proposed by Saito and Fuku\-yama 
\cite{Saito97,Saito99}. A phase Hamiltonian 
that describes a  one-dimensional AF chain
coupled with a 3D lattice distorsion field was used.
The observation of both SP and AF signatures 
can be consequently explained by considering a ground state
where two long range order parameters
coexist with a spatial variation. 
The lattice dimerisation is minimal near the doping centers,
where the staggered magnetisation takes its maximal value.
Moreover, the phase transition from D-AF to U-AF is predicted to 
be of first order if the spin-phonon coupling is small when compared 
with the interchain interaction. A first experimental evidence
of the spatial inhomogeneity for the magnetic moments was seen 
with muon spin rotation experiments on Si- and Zn-doped 
CuGeO$_{3}$ compounds \cite{Kojima97}.
    
In the present work we report a detailed study of the
temperature-concentration (T-x)
phase diagram of Si-\-doped CuGeO$_{3}$ single crystals by means of
neutron diffraction as well as zero-field field
muon spin rotation ($\mu$SR).
The different signals observed with the neutrons,
namely the AF and SP superlattice peaks,
are interpreted in terms of volume fractions
via the analysis of the  zero-field muon spectra.
%Additional measurements performed with an external field 
%transverse to the initial muon polarisation
%provide information about the dimension of the 
%regions that contribute to the different signals.
Finally, the local magnetic moment measured by $\mu$SR 
is compared with the macroscopic order parameter obtained
by neutron diffraction.

\section{Experimental Details}\label{sec:exp-det}

The Si-doped CuGeO$_{3}$ single crystals have been 
grown using the floating zone method.
The impurity concentration has been 
determined by means of inductively coupled plasma atomic emission 
spectroscopy (ICP-AES), with an accuracy of about 0.1\%.
The samples have been characterised by bulk susceptibility
measurements which have been published elsewhere \cite{Masuda00-2}.

The elastic neutron scattering experiments have been performed on the 
three-axis spectrometer for cold neutrons TASP, at the neutron 
spallation source SINQ, on a series of doped
CuGe$_{1-x}$Si$_x$O$_3$ crystals (0.7\% $ \leq x \leq$ 3.8\%).
The incident neutron wave vector was kept 
fixed at $k_{i}$ = 2.662 \AA$^{-1}$. Higher-order neutrons have 
been suppressed by using a pyrolitic graphite filter.
The samples have been oriented with the ($0kl$) zone in the 
scattering plane,
and mounted in an ILL-type cryostat, which achieves
a base temperature of 1.5K.

Zero field  $\mu$SR measurements have been performed  
on the spectrometers GPS and LTF, at the muon facility of PSI.
The polarisation of the incident muon beam
was parallel to the a-axis of the sample. 
%In transverse field 
%experiments, the external field was applied perpendicular to the
%initial muon polarisation.

%\subsection{Subsection title}
%\label{sec:2}

\section{Results}\label{sec:results}

\subsection{Neutron diffraction}\label{sec:nd}

\begin{figure}
    \resizebox{0.5\textwidth}{!}{%
     \includegraphics{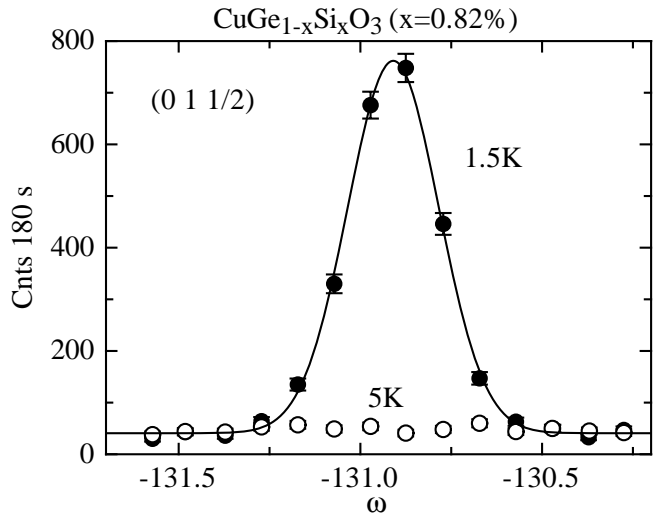}
    }
    \caption{ Profile for the magnetic superlattice peak (0 1 1/2) 
     for the x=0.82\%-doped sample, at 1.5K. A similar 
    measurement at 5K shows the absence of higher order contamination.
    }
    \label{fig:af-profile}
\end{figure}

\begin{figure}
    \resizebox{0.5\textwidth}{!}{%
     \includegraphics{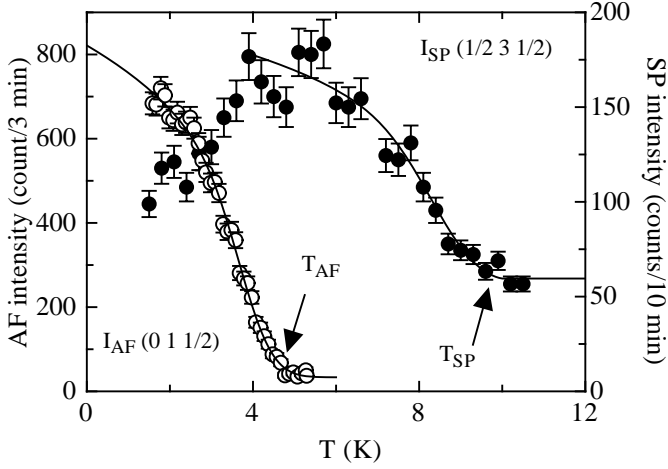}
    }
    \caption{Temperature dependence of the (0~1~1/2) 
    antiferromagnetic and (1/2~3~1/2) spin-Peierls superlattice peaks in 
    the 0.82\% Si-doped CuGeO$_{3}$ single crystal.
    The solid line for (0~1~1/2) results from a fit according
    to Eq.~\ref{eq:maxint}. A similar function was used for the SP peak.
    }
    \label{fig:tn-tsp}
\end{figure}

A typical profile of an antiferromagnetic superlattice peak
is shown in Fig.~\ref{fig:af-profile} for the 0.82\% sample.
The temperature dependence of the maximum intensity
is presented in Fig.~\ref{fig:tn-tsp}.
In order to confirm the onset of the SP dimerisation in 
the low doping regime, the reflection (1/2 3 1/2) was also measured.
The magnetic intensity has been fitted with 
the following equation \cite{Hase96}: 

\begin{equation}
	I(T) = N \int_{T}^{\infty} ({T_{N}' - T \over T_{N}'})^{2\beta} 
	\exp\Big[-{(T_{N}'-T_{N})^{2}\over 2\Delta T_{N}^{2}}\Big] dT_{N}' + BG
	\label{eq:maxint}
\end{equation}

\noindent which describes the critical behaviour of the magnetic intensity 
near the AF transition, weighted with a Gaussian distribution
for T$_{N}$ that accounts for inhomogeneities in the sample.
N is a normalisation constant, $\beta$ the critical exponent 
and $BG$ the nonmagnetic background contribution. 
The exponent $\beta$ 
remains almost constant over the whole diagram, with an average 
value of about $0.20 \pm 0.04$, similar to that observed in Zn-doped
crystals \cite{Hase96}. The inhomogeneity in the concentration
provides a $\Delta T_N$ of about $0.3 \pm 0.05$K.
The temperature dependence of the SP peak shown in Fig.~\ref{fig:af-profile} was 
fitted with a function similar \cite{Nakao99} to the one of Eq.~\ref{eq:maxint}.

The effective magnetic moments $\mu_{eff}$ have been calculated by normalising 
the magnetic intensity at the saturation value
with the nuclear structure factor of the (0~2~1) 
reflection, and corrected for the magnetic form factor of the free
Cu$^{2+}$ ion \cite{ITC}. We emphasise that the
neutron diffraction technique is a non local probe, and hence 
does provide a macroscopic order parameter, that is averaged
over the whole volume of the sample.

\begin{figure}
    \resizebox{0.5\textwidth}{!}{%
     \includegraphics{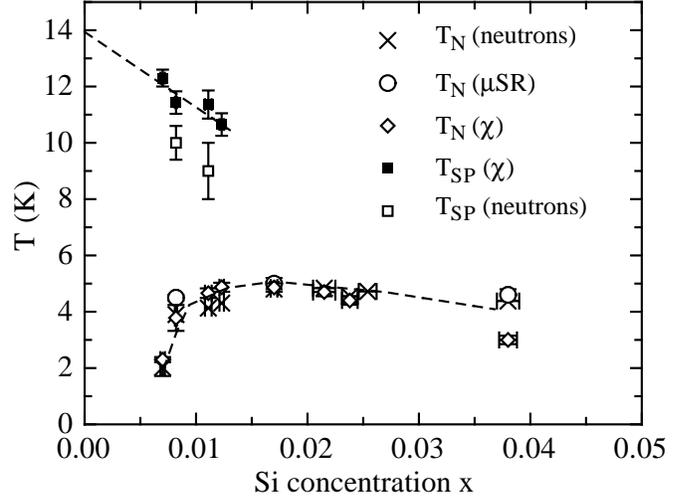}
    }
    \caption{Temperature-concentration phase diagram of the Si-doped CuGeO$_3$.
    The AF ordering temperature T$_N$ and the spin-Peierls transition 
    temperature T$_{SP}$
    have been determined by bulk susceptibility \cite{Masuda00-2} and neutron diffraction
    experiments. The T$_N$ determined by $\mu$SR measurements are also
    reported. The dashed lines are a guide to the eye.
    }
    \label{fig:t-x}
\end{figure}

As a next step, selected samples (0.82\%, 1.7\%, 2.38\% and 3.8\%)
belonging to the series of single crystals
measured with neutrons 
have been investigated by muon spin rotation spectroscopy
in the three temperature regions
that have been determined with neutron and susceptibility
measurements, according to the experimental
T-x diagram presented in Fig.~\ref{fig:t-x}.
The values obtained from the various techniques are found to be in 
good agreement, indicating that the muon and the neutrons observe a magnetic 
ordering at the same transition temperature.
The observation of the SP dimerisation in susceptibility 
data for the samples in the low 
doping regime is also confirmed by neutrons. The discrepancy
between both methods is due to the fact that the SP superlattice
peak that is measured with neutrons becomes extremely small
while approaching the critical concentration.
Furthermore, we emphasise that T$_{SP}$ is not observed above x=1.2\%.

\subsection{$\mu$SR results}

In the paramagnetic phase, the zero-field muon spectra have been 
fitted
with the Kubo-Toyabe function which accounts for the muon 
depolarisation originating from the nuclear moments (the 
depolarisation rate is of the order of 0.08 MHz).
By lowering the temperature, in the low doping regime, the muon 
spectrum can be
still explained by the same relaxation function, as it was 
observed for pure CuGeO$_3$ \cite{Lappas94}.
In the magnetically ordered phase, below T$_N$, the muon 
depolarisation $P_{\mu}(t)$
in zero field was analysed over the whole doping regime 
with the following function (see also \cite{Kojima97}):
    
\begin{eqnarray}    \label{eq:muon-depol}
    P_{\mu}(t)& = & A_{rlx} \exp(-\Delta t) \\ \nonumber
              & + & A_{osc} \exp(-\Gamma t)\cdot\cos(2\pi\nu t + 
\phi).
\end{eqnarray}   

The first term in Eq.~\ref{eq:muon-depol} describes a non-precessing 
part of the muon signal, which relaxes at a rate $\Delta$, whereas 
the second term reflects a precessing part of the muon polarisation, 
where $\Gamma$ is the depolarisation rate and $\phi$ the initial 
phase shift. The total observed amplitude ($A_{rlx} + A_{osc}$) is 
significantly lower than the total amplitude observed above T$_N$, 
indicating that part of the muon ensemble is depolarised within the 
dead time of the spectrometer ($i.e.$ a depolarisation rate higher 
than ca. 150~MHz).

According to the T-x diagram shown in Fig.~\ref{fig:t-x},
the emergence at low temperature of the precessing signal
($A_{osc}$) in zero-field is attributed to the
ordering of the Cu$^{2+}$ moments near the doping centers.
In this vein, the frequency $\nu$, which is proportional to the 
magnetic field $B_{\mu}$ at the muon site, mirrors the static 
ordering of the moments.

\begin{figure}
    \resizebox{0.5\textwidth}{!}{%
     \includegraphics{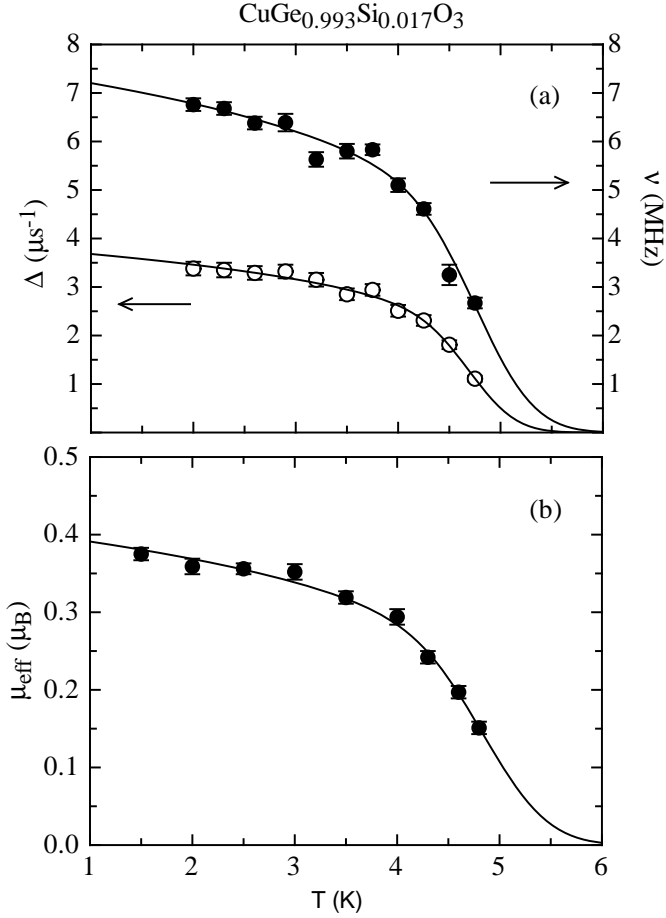}
    }
    \caption{Temperature dependence of the relaxation rate $\Delta$
    and the precessing frequency $\nu$ for the 1.7\% doped single 
crystal 
    measured in zero-field $\mu$SR (a).
    The magnetic moment measured by means of neutron diffraction for 
the same
    sample is shown in (b). The solid lines are fits according to 
    Eq.~\ref{eq:maxint}.
    }
    \label{fig:mueff-delta-nu}
\end{figure}

\begin{figure}
    \resizebox{0.5\textwidth}{!}{%
     \includegraphics{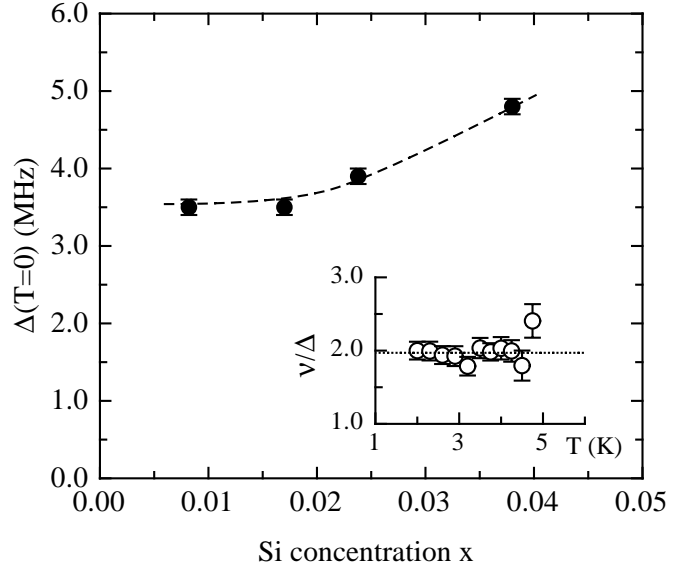}
    }
    \caption{Muon relaxation rate $\Delta$  extrapolated to T=0, 
    plotted as a function of doping. The value is proportional 
    to the local magnetic moment at the muon site (see inset). 
    The dashed line is a guide to the eyes. The inset 
    shows the ratio $\nu/\Delta$ for the x=1.7\% sample, confirming
    the proportionality between both parameters ($\nu/\Delta \approx$ 
2, 
    as indicated by the dotted line).
    }
    \label{fig:freq}
\end{figure}

The relaxing component in Eq.~\ref{eq:muon-depol} is ascribed to 
sample regions where no coherent static magnetism is present.
 However, the similar 
temperature dependence of $\Delta$ and $\nu$ (see 
Fig.~\ref{fig:mueff-delta-nu} and the inset of Fig.~\ref{fig:freq}) 
indicates that these regions have reduced dimensions, leading to a 
detectable influence of the ordered neighbouring regions on the field 
distribution $\Delta/\gamma_{\mu}$ (where $\gamma_{\mu}$ is the 
gyromagnetic ratio of the muon) at the muon sites. This supports the 
idea that there is no {\em macroscopic} phase separation in the 
samples over the whole doping range, similarly to what 
was observed by Kojima et al. for Si- and Zn-doped CuGeO$_3$
compounds in the low doping regime \cite{Kojima97}. Therefore, and 
similarly to the parameter $\nu$, the parameter $\Delta$ can be taken 
as a measure of the static ordered moments. Since $A_{rlx} \gg 
A_{osc}$ the parameter $\Delta$ appears to be better determined than 
$\nu$ and consequently $\Delta$ will be utilised in the following 
discussion as a measure of the static moment.

The doping dependence of the relaxation rate $\Delta$, 
extrapolated at the saturation value, is shown in Fig.~\ref{fig:freq}.
The dependence upon Si-doping is almost constant within 
the precision of the experimental values.
$\Delta$ exhibits however an increase above x=2.38\%.
According to the direct relation between $\Delta$ and $\nu$,
this behaviour indicates that 
within the doping range that has been investigated
the local magnetic field at the muon sites, and therefore the ordered 
moment increases only slightly with increasing temperature.

\begin{figure}
    \resizebox{0.5\textwidth}{!}{%
     \includegraphics{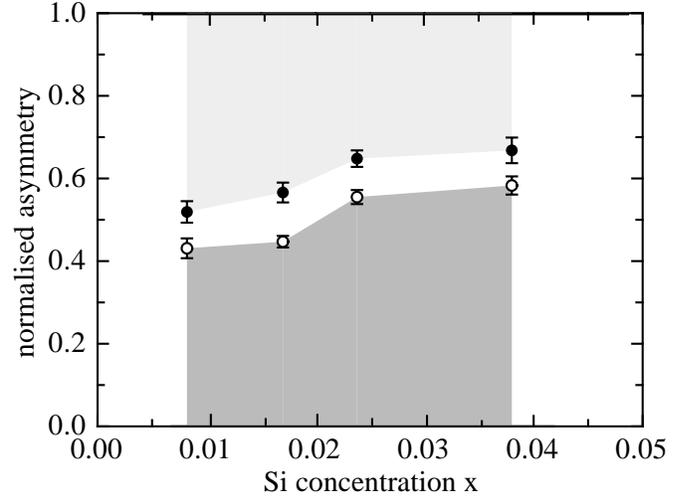}
    }
    \caption{
    The normalised amplitudes of the muon depolarisation 
    $\hat{A}_{rlx}$ (light grey area)
    and $\hat{A}_{osc}$ (white area), obtained from 
    Eq.~\ref{eq:muon-depol}, as a function of Si doping. The 
quantities
    have been normalised to the total muon asymmetry $A_{tot}$.
    The dark grey region refers to lost part of the muon signal 
    (i.e. 1 - $\hat{A}_{rlx}$ - $\hat{A}_{osc}$).
    }
    \label{fig:musr-ampl}
\end{figure}

Figure~\ref{fig:musr-ampl} shows the evolution
of the the normalised amplitudes $\hat{A}_{rlx}$ (light grey area)
and $\hat{A}_{osc}$ (white area), as a function of Si concentration.
The amplitudes have been normalised with respect to the total 
muon asymmetry $A_{tot}$, that was determined in the paramagnetic 
phase.
We obtain that the amplitude corresponding to the magnetically 
ordered phase,
$\hat{A}_{osc}$, represents about 10\% of $A_{tot}$. The relaxing 
part $\hat{A}_{rlx}$ lies on the other hand between 40\% and 60\%.
The part of $A_{tot}$ (dark grey area in Fig.~\ref{fig:musr-ampl})
that could not be detected 
($\approx$ 40\%) will be discussed below.

%Now we combine musr and neutrons
In order to compare the local magnetic properties of the doped
samples, as provided by the muons,
with the macroscopic order parameter measured with neutrons, 
we use the quantity $\Delta\cdot\hat{A}_{osc}$.
In this product, the local magnetic moment (proportional to $\Delta$) 
is hence weighted with the volume fraction ($\hat{A}_{osc}$) of 
magnetised regions present in the sample. 

\begin{figure}
    \resizebox{0.5\textwidth}{!}{%
     \includegraphics{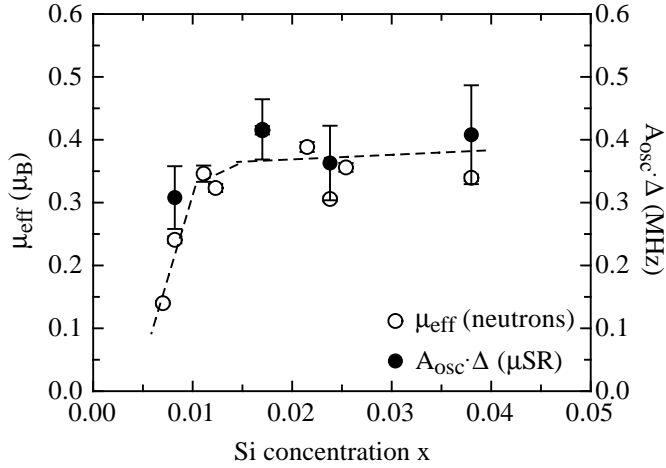}
    }
    \caption{The order parameter $\mu_{eff}$
    as a function of Si doping $x$ measured with the 
    neutron diffraction method (open circles). The effective magnetic 
    moment determined by $\mu$SR (explained in the text) 
    is also shown (solid circles). The dashed line is a 
    guide to the eyes.}
    \label{fig:mu-both}
\end{figure}

The dependence of the AF order parameter $\mu_{eff}$ 
as a function of concentration is shown
in Fig.~\ref{fig:mu-both}, as obtained 
by both neutron diffraction and zero-field $\mu$SR methods.
$\mu_{eff}$ is seen to increase with the  doping concentration, 
reaches a maximum at x$_{c}$= 1.7\%, where the SP signal
is not observed anymore (see Fig.~\ref{fig:t-x}).

\section{Discussion}\label{sec:disc}

The agreement between neutron and $\mu$SR results (see 
Figs.~\ref{fig:t-x} and \ref{fig:mu-both})
 allows to associate the precessing signal, referred to
A$_{osc}$, with the D-AF phase in the low doping regime,
and with the U-AF phase in the high doping regime.

The purely relaxing signal, A$_{rlx}$, is attributed,
 for $x \leq$ 1.23\%, to the regions where the 
lattice dimerisation is maximal.
However, for higher doping concentrations,
this amplitude does not disappear.
It is thought to arise from an increase of static 
magnetic disorder that is induced by high doping.

As the sum of $\hat{A}_{rlx}$ and $\hat{A}_{osc}$ constitutes not 
more than
60\% of the total muon asymmetry in the whole investigated doping 
regime,
we may attribute the loss of signal to the boundary between 
the AF regions and the dimerised regions
where a rapid spatial variation of the magnetisation gives rise to
large field gradient and therefore leads to 
a depolarisation rate too fast to be observed by the $\mu$SR 
spectrometer. However, the neutrons are sensitive to these 
boundary regions that contribute to the intensity of the AF peaks.

The $\mu$SR measurements have shown on the other hand
that the muon signal described by A$_{rlx}$ is influenced by the
dipolar fields of the magnetically ordered regions 
and still undergoes a depolarisation.
This is a clear sign that the domains contributing to the purely
relaxing muon signal have a limited size. 
Considering the magnitude of the magnetic moments that are determined
by neutron diffraction, 0.1 to 0.4 $\mu_B$,
the maximal distance over which the muon still sees
a dipolar field is estimated to be $\approx$ 15 \AA. 
By considering an homogeneous distribution of impurities along the 
spin chains, this distance corresponds to $\approx$ 5 lattice units
in both directions on each side of the muon site. 
%%%%% Kojima ... 
The present measurements confirm the spatial inhomogeneity
of the magnetic moments that was predicted theoretically 
\cite{Saito97,Saito99}.

The maximum in the order parameter $\mu_{eff}$ that was
observed by neutron diffraction at x$_c$=1.7\% refers to a 
macroscopic value.
In other words, the magnetic moment determined by this technique
is averaged over the whole volume of the sample. 
The results provided by $\mu$SR indicate that the volume fraction 
of the magnetically ordered regions as well as the boundary
regions varies upon doping.
On the other hand, no anomaly is seen in the local magnetic moment
near the critical value x$_c$.
We suggest therefore that the variation of the macroscopic AF order 
parameter
upon doping is controlled by a variation in the volume fraction of 
magnetic
domains rather than by a drastic change of the magnetic moments.

\section{Conclusion}

The compilation of information provided by neutron and $\mu$SR
have provided a detailed insight about the origin of
the various magnetic and non magnetic contributions
observed in the Si-doped CuGeO$_{3}$.
The $\mu$SR results have revealed that the 
regions that contribute to the antiferromagnetic
and lattice dimerisation peaks observed by neutron and susceptibility
in the Si-doped CuGeO$_3$ have a typical size of 5 lattice units
along the spin chains.

Furthermore, the doping variation of the magnetic moment measured 
with neutrons could be explained via the evolution
of the volume fraction attributed to magnetically ordered
regions near the impurity centers.

\vspace {0.5 cm}
One of the authors (F.S.) would like to thank the ETH-Council
for the supply of a research grant.
The U.S. Department of Energy
is also acknowledged for financial support (G.S.).

\end{document}